\documentstyle[aps,prl,twocolumn]{revtex}
\input epsf

\begin{document}
\draft
\title{Observation of harmonic generation and nonlinear coupling in the collective dynamics of a Bose condensate}
\author{G. Hechenblaikner, O.M. Marag\`o, E. Hodby, J. Arlt, S. Hopkins\footnote{Present address:
SCOAP, CPES, University of Sussex, Falmer, Brighton BN1 9QH}, and C.J. Foot }
\address{Clarendon Laboratory, Department of Physics, University of Oxford,\\
Parks Road, Oxford, OX1 3PU, \\
United Kingdom.}
\date{\today}

\maketitle

\begin{abstract}
We report the observation of harmonic generation and strong nonlinear coupling of two collective modes of a condensed gas of rubidium atoms.
Using a modified TOP trap we changed the trap anisotropy to a value
where the frequency of the $m=0$ high-lying mode corresponds to twice the frequency of the $m=0$ 
low-lying mode, thus leading to strong nonlinear coupling between these modes. 
By changing the anisotropy of the trap and exciting the low-lying mode
we observed significant frequency shifts of this fundamental mode and also the generation of its second harmonic.
\end{abstract}

Ever since the  first Bose condensed gases were produced \cite{Anderson},
collective excitations have played a key role in the theoretical models and
their experimental verification.
Measurements of collective excitation frequencies may be compared
with theoretical predictions and the first measurements of the lowest collective
modes of a Bose gas were carried out soon after the first condensates were made \cite{jin,mewes}.
Those  experiments verified that the Gross-Pitaevskii (GP) equation, also called the nonlinear 
Schr\"odinger equation (NLSE), gives a very accurate prediction of the frequencies of the measured collective modes.
The nonlinearity of the condensate is manifested as a term in the GP equation proportional to the mean-field
or condensate density. In this experiment we have observed effects arising directly from this term. 

The spectroscopy of the excited states of the condensate has been extended to include measurements of
damping rates and frequency shifts as a function of temperature (in the range $0<T/T_c<1$, where $T_c$ is
the critical temperature of condensation)\cite{jin2}.
The finite temperature measurements test dynamical aspects of the theory which require higher order terms
in the calculations and as yet there is not full agreement \cite{hutchinson}.

Recent theoretical work has contributed to a better understanding of damping of the collective excitations
at finite temperature \cite{damping}.
Although mechanisms that lead to damping are complex one may consider two broad types of behavior.
Firstly, energy may be lost from a mode by being transferred into a multitude of other modes
through the nonlinear coupling that is intrinsic to condensates.
The second cause of damping is scattering of non-condensate particles in so-called Landau processes.
The relative importance of these two processes depends on the shape of the condensate. In this work we
have been able to vary the aspect ratio of the condensate around a value where nonlinear coupling is large and hence likely to dominate over other processes.

In the hydrodynamic limit which applies to condensates with a large number of atoms, the
GP equation reduces to the hydrodynamic equations of nondissipative fluid dynamics.
In this limit Stringari \cite{stringari} found analytic solutions for the frequencies of the collective modes
and Edwards et al.\cite{edwards} extended the calculations to the non-hydrodynamic region by direct integration of the
GP equation.
The measured oscillation frequencies correspond to those obtained from linear response theory only 
in the limit of small driving and response amplitudes (zero amplitude limit). 
For strong driving the inherent nonlinearity 
in the condensate becomes apparent and the actual response frequency is shifted from the zero amplitude limit, 
$\omega_0$.
The shift $\Delta\omega$ is predicted to be proportional to the square of the 
response amplitude to first order approximation, i.e.
$\Delta\omega/\omega_0 =A^2\delta(\lambda)$, where $A$ is the excitation
response amplitude, $\delta(\lambda)$ is a nonlinearity factor which varies for 
different trap geometries \cite{dalfovo} characterized by $\lambda=\omega_z/\omega_r$
(the ratio of axial to radial trap frequencies). 
Thus a frequency shift of the collective excitation can be achieved in two ways: either by increasing the 
driving and consequently the response amplitude of the oscillation or by changing the trap geometry and hence the
nonlinearity factor. In previous work only the amplitude dependence of the frequency shift for a fixed trap geometry 
has been studied \cite{jin,mewes,jin2}.
This paper describes the use of a trap with adjustable anisotropy \cite{ensher}
 to study the effects of trap geometry on collective excitations. 

\begin{figure}
\begin{center}\mbox{ \epsfxsize 3.2 in\epsfbox{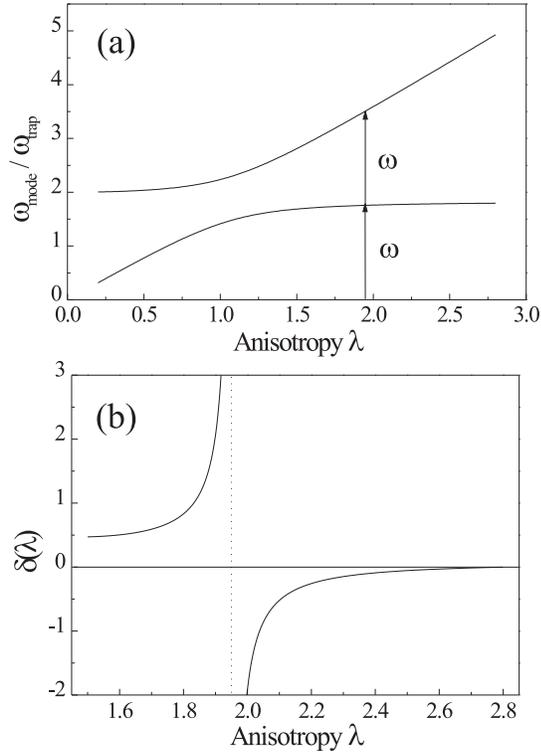}}\end{center}
\caption{(a) The coupling between the low and the high-lying mode is strongest for a trap anisotropy of $\lambda=1.95$.
In this geometry the frequency of the high-lying mode is twice the frequency of the low-lying mode.
(b) The nonlinearity factor $\delta(\lambda)$ is plotted against the anisotropy $\lambda$. 
In this region strong nonlinear features are observed around the resonance for second harmonic generation. 
Both plots are based on the hydrodynamic theory.}
\label{resonance2}
\end{figure}

The three lowest collective modes of a condensate in an axially symmetric trap ($\omega_x=\omega_y=\omega_r$) are the $m=2$ mode, the
low-lying $m=0$ mode and the high-lying $m=0$ mode, where $m$ is the azimuthal angular momentum  
(the trivial centre of mass motion is not considered).
The $m=2$ mode corresponds to a quadrupole type excitation in the radial plane.
The low-lying $m=0$ mode corresponds to a radial oscillation of the width which is 
out of phase with an oscillation along the trap axis.
The high-lying $m=0$ mode is an in-phase compressional mode along all directions (breathing mode).
In our experiment we changed the trap anisotropy parameter $\lambda$ around a resonance between
the low-lying and the high-lying $m=0$ modes. At this resonance two quanta of excitation of the low-lying mode
are converted into one quantum of excitation of the high-lying mode, i.e. the second harmonic
of the low-lying collective mode is excited. The theoretical plot (for the hydrodynamic
limit) in Fig.\ref{resonance2}a shows that at a trap anisotropy of $\lambda\approx 1.95$
 (resonance) the frequency of the 
high-lying $m=0$ mode corresponds to exactly twice the frequency of the low-lying mode.\\
 The nonlinearity factor $\delta(\lambda)$ also changes dramatically across the resonance.
A theoretical plot of $\delta(\lambda)$ against $\lambda$ is given in Fig.\ref{resonance2}b 
using the formula given in \cite{dalfovo}.

Our experimental apparatus for creating Bose condensates is described elsewhere \cite{arlt} and
we only briefly mention the relevant features here. Atoms are loaded into a TOP trap 
and then  evaporatively cooled to the quantum degenerate regime. We cool the atoms to 
well below the critical temperature where no thermal cloud component is observable ($\sim 0.5 T_c$).
For this experiment the usual TOP trap configuration with a rotating field in the radial plane
was modified by the addition of a pair of 'z-coils' in Helmholtz configuration 
which produced an oscillating axial bias field along the (vertical) quadrupole coil axis. 
The total field in the trap is 
\begin{eqnarray}
{\mathbf B}(t)=B'_q\left(x{\mathbf e}_{x}+y{\mathbf e}_{y}-2z{\mathbf e}_{z}\right)+\nonumber\\
B_r\left(\cos \omega_t t\ {\mathbf e}_{x}+\sin \omega_t t\ {\mathbf e}_{y} \right)+
B_z\cos 2\omega_t t\ {\mathbf e}_{z}.
\label{topfield}
\end{eqnarray}
The first term is the static quadrupole field and is written in terms of its gradient
$B'_q$ in the radial direction. The second term is the conventional TOP field of magnitude $B_r$ 
rotating in the xy-plane at a frequency $\omega_t$. 
The third term is the additional z-bias field  of magnitude $B_z$ modulated at twice 
the frequency of the field in the xy-plane.
Hence the locus of the quadrupole field does not  simply follow a planar circular path 
like that of a standard TOP trap but it also moves up and down following a saddle shape.

The time average of this field configuration yields the value for a trap frequency ratio 
that is smaller than $\lambda=\sqrt 8$ in a TOP trap with $B_z=0$.
By increasing $B_z$ the anisotropy $\lambda$ is tuned
continuously from $2.83$ to $1.6$. The practical limit on $\lambda$ in our present apparatus arises 
because of the noise created by the amplifier driving the z-coils 
which induces Zeeman substate changing  transitions for large amplitudes
of $B_z$.
This limit may be overcome to create a spherical trap ($\lambda=1$) as described in \cite{hodby},
which also gives more details of the technique.
We calibrated the trap by measuring the centre of mass oscillations of a thermal cloud for various amplitudes
of $B_z$. The measured frequencies agree to better than $1\%$ with a theoretical prediction
obtained by numerically time-averaging the field in Eq.\ref{topfield}, i.e. integrating
over one complete TOP cycle. 
Note that the radial trap frequency remains approximately constant 
whereas the axial frequency decreases as the amplitude $B_z$ increases.
The trap frequencies for this experiment were $126$ Hz radially and the axial frequencies varied from
$356$ Hz at $B_z/B_r=0$ ($\lambda=\sqrt 8$) to $194$ Hz at $B_z/B_r=2.3$ ($\lambda=1.6$).

Once the trap had been calibrated  we investigated the behavior of the $m=0$ mode for various anisotropies.
To excite the $m=0$ low-lying mode, the TOP-field amplitude
$B_r$ was modulated sinusoidally at a frequency of $225$ Hz (matching the hydrodynamic mode frequency
in a $126$ Hz radial trap with $B_z=0$) for a period of $8$ cycles ($35$ ms). 
Note that the actual measured mode frequency for the above trap parameters differs from $225$ Hz
because the finite number of atoms leads to deviations from the hydrodynamic limit. 
Also, the change of mode frequency with trap anisotropy (see Fig.\ref{resonance2}a) means that
the driving frequency does not exactly match the hydrodynamic mode frequency for traps with 
anisotropies different from $2.83$ (the value for $B_z=0$), but it is close enough
to excite the fundamental mode for all our measurements.

After being excited the condensate was left to oscillate for variable times of between $2$ to $30$ ms before
the magnetic trap was switched off and the condensate allowed to expand freely for $12$ ms to make a
 time-of-flight (TOF) measurement.
Then absorption images were taken of the condensate from which we extracted both the radial and axial widths
and the total number of atoms in the condensate (typically $2\times 10^4$).
The radial TOP field produced an $8\%$  modulation of the trap spring constant and 
we measured response amplitudes of around $20\%$ in TOF. 
These correspond to even smaller amplitudes of oscillation for the condensate before expansion 
which produce no measurable frequency shift  in the ordinary TOP trap (for $\lambda=2.83$ 
the nonlinear coupling $\delta$ for the $m=0$ mode is
close to zero, see Fig.\ref{resonance2}b). 
Even for strong driving and large response amplitudes no significant frequency shift of the $m=0$ mode 
from the zero amplitude limit has been observed before \cite{jin2}.

\begin{figure}
\begin{center}\mbox{ \epsfxsize 3.2 in\epsfbox{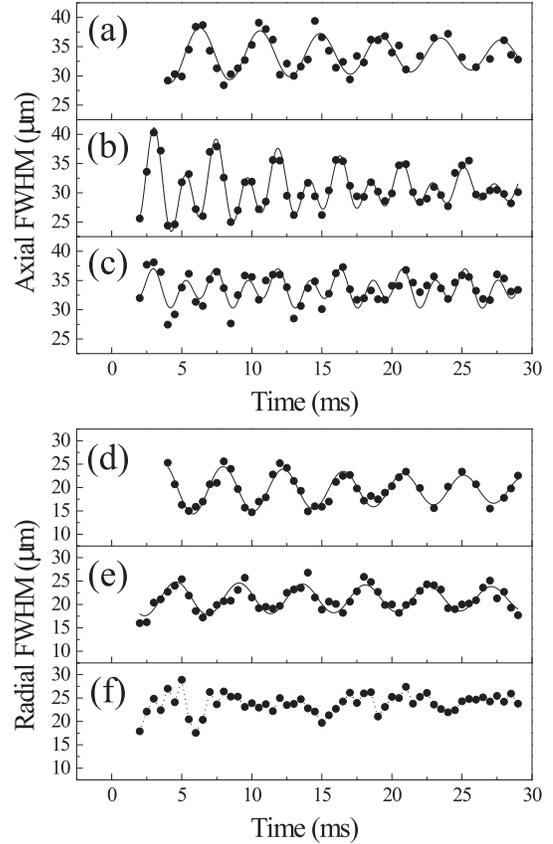}}\end{center}
\caption{ Shape oscillations of the BEC are plotted against time for various geometries.
Sub-plots (a,b,c)  show oscillations in the axial width and sub-plots (d,e,f) 
oscillations in the corresponding radial width.
(a) Away from resonance ($\lambda=2.64$) only a single frequency appears. 
(b) Close to resonance  ($\lambda=1.97$) there is a strong admixture
of a second harmonic contribution. 
(c) On resonance ($\lambda=1.93$) the second harmonic contribution dominates.
(d,e) The radial widths oscillate at the frequency of the fundamental mode ($m=0$ low-lying mode)
for $\lambda=2.64,1.97$.
(f) The oscillation of the fundamental mode disappears on resonance.}
\label{data}
\end{figure}

\begin{figure}
\begin{center}\mbox{ \epsfxsize 3.2 in\epsfbox{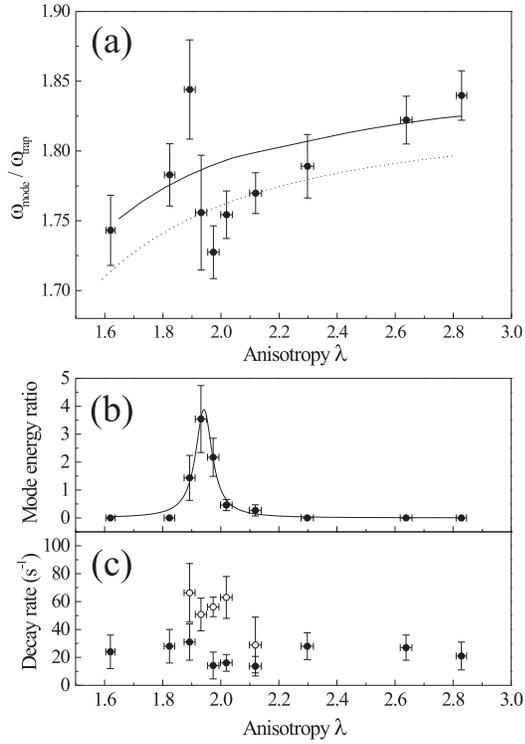}}\end{center}
\caption{(a) The frequency of the $m=0$ low-lying mode in units of the radial trap frequency as a function of
the trap anisotropy for various geometries. The dotted line is the prediction of the hydrodynamic theory
and the solid line was calculated from the GP equation.
 (b) The measured energy ratio of the higher to the lower-lying 
mode is plotted against the trap anisotropy. The solid line is a Lorentzian fit to the data.
 (c) The damping rates for the fundamental (black circles) and the second harmonic (white circles) frequency 
are plotted against the anisotropy $\lambda$.} 
\label{freq}
\end{figure}

The anisotropy of our trap was tuned across the resonance between the low and the high-lying $m=0$ mode. 
We observed the occurrence of the second harmonic frequency when
we tuned our trap close to resonance (Fig.\ref{data}b) and on resonance it was the dominant component (Fig.\ref{data}c).
It arises from the excitation due to nonlinear coupling of the $m=0$ high-lying mode
and could only be observed in the oscillations of the axial width.
The radial width oscillated at a single frequency corresponding to the $m=0$ low-lying mode and
did not have a second frequency component (Fig.\ref{data}d,e).
This is also what we found in simulations based on the hydrodynamic equations and indicates that the geometries
of the low and high-lying modes are such that the excitation of the high-lying mode can only be observed
by measuring the axial widths. We obtained the low-lying mode frequency from a single frequency fit
to the radial oscillations and also from a two frequency fit to the axial oscillations and 
found good agreement between the two values (exponential decay factors were included in both fits).
The frequency of the second harmonic (corresponding to the high-lying mode) was obtained as 
the second frequency component in the fit along the axial direction.
We found that the fundamental frequency was strongly suppressed on resonance 
in both the radial and axial oscillations as shown in Fig.\ref{data}c,f. This indicates that
the low-lying mode is not populated anymore and all the energy of the excitation has been transferred to the high-lying 
mode \cite{morgan}. The large error for the point at resonance ($\lambda=1.93$) in Fig.\ref{freq} reflects the difficulty in obtaining
the fundamental component from the data in Fig.\ref{data}c.

The frequencies of the fundamental mode for various trap geometries were normalized by the
corresponding radial trap frequency and plotted as a function of 
the trap anisotropy parameter $\lambda$ in Fig.\ref{freq}a. 

The dotted line is the prediction of the small amplitude
mode frequencies in the hydrodynamic limit as given in \cite{stringari}. The solid line was calculated
by Hutchinson \cite{hutchinson2} from the GP equation for the atomic number and trap frequencies of our experiment.
Figure \ref{freq}a shows that the measured mode frequency is above the hydrodynamic  prediction for the ordinary
TOP trap with $\lambda=2.83$ in agreement with previous experiments \cite{jin}.
Far from resonance the data agree very well with the finite number prediction (solid line). 
However, when approaching the resonance from above the nonlinear coupling becomes large enough to shift
the mode frequency below the predicted curve.
On the lower  ($\lambda<1.95$) side of the resonance the frequency shift becomes positive in agreement 
with the predicition illustrated in Fig.\ref{resonance2}b.
One can see that the measured frequency shifts follow a dispersive curve 
which crosses zero at $\lambda=1.93\pm 0.02$, in good agreement with the prediction
of $\lambda=1.95$ for the hydrodynamic regime \cite{dalfovo}. 
Figure \ref{freq}b shows a plot of the ratio of the squared amplitudes  (higher to lower-lying mode) 
which is proportional to the ratio of the mode energies.
Away from resonance we observed no second harmonic contribution and the ratio (and its error) is taken as zero.
These data have been fitted by a simple Lorentzian centered at $\lambda=1.94\pm 0.02$ with HWHM$=0.035\pm 0.005$.
The damping rates are plotted against the anisotropy $\lambda$ in Fig. \ref{freq}c.
They were around $20 s^{-1}$ for the fundamental and $50 s^{-1}$ for the second harmonic frequency. 

In conclusion, we have a TOP trap in which the anisotropy can be tuned 
whilst keeping the trap frequencies high \cite{ensher}. 
This offers the opportunity to investigate a range of interesting phenomena in Bose-Einstein 
condensates related to the trap geometry, in particular the greatly enhanced nonlinear effects 
for certain anisotropy parameters $\lambda$. 
We observed the generation of the second harmonics of the $m=0$ low-lying mode
frequency as well as a dispersive resonance in the mode frequency at $\lambda=1.93\pm 0.02$.
Future improvements and modifications  of our experiment will allow us to study another predicted resonance 
at $\lambda=1.5$ as well as additional nonlinear phenomena such as the collapse and revival of oscillations and 
the onset of chaos for stronger driving amplitude \cite{salasnich}.
A particularly interesting case is the spherical trap ($\lambda=1$) where all frequencies are degenerate.
Quantitative theoretical predictions have been made for the damping rates in a spherical trap \cite{sphere} which can
be compared with the experiment to provide a stringent test.

We would like to thank all the members of the Oxford theoretical BEC group,
in particular D. Hutchinson for providing the theoretical prediction, K. Burnett, M. Davis, S. Morgan and M. Rusch 
for their help and many useful discussions.

This work was supported by the EPSRC and the TMR program (No. ERB
FMRX-CT96-0002). O.M. Marag\`{o} acknowledges the support of a
Marie Curie Fellowship, TMR program (No. ERB FMBI-CT98-3077).


\end{document}